\documentclass[]{aa} 
\begin{document}
\def\rchi{{${\chi}_{\nu}^{2}$}}
\newcommand{\pcmsq} {cm$^{-2}$}
\newcommand{\xmm} {\sl XMM-Newton}
\newcommand{\ergscm} {erg s$^{-1}$ cm$^{-2}$}
\newcommand{\ergss} {erg s$^{-1}$}
\newcommand{\ergsd} {erg s$^{-1}$ $d^{2}_{100}$}
%
\title{XMM-Newton observations of the dwarf nova YZ Cnc in quiescence}
\titlerunning{XMM-Newton observations of YZ Cnc}
\authorrunning{Hakala et al}


\author{Pasi Hakala\inst{1}, Gavin Ramsay \inst{2}, Peter Wheatley
\inst{3}, Emilios Harlaftis \inst{4},
\and C. Papadimitriou \inst{5}
}

\offprints{P. Hakala}

\institute{$^{1}$Observatory, P.O. Box 14, FIN-00014 University of
Helsinki,
Finland\\
$^{2}$Mullard Space Science Lab, University College London,
Holmbury St. Mary, Dorking, Surrey, RH5 6NT, UK\\
$^{3}$Department of Physics \& Astronomy, University of Leicester,
University Road, Leicester, LE1 7RH, UK\\
$^{4}$Institute of Space Applications and Remote Sensing, National
Observatory of Athens, P.O. Box 20048, Athens - 11810, Greece\\
$^{5}$Institute of Astronomy \& Astrophysics, National Observatory of Athens,
P.O. Box 20048, Greece.\\
}


\date{}

\abstract{

We present results from the {\xmm} observations of the dwarf nova YZ Cnc 
in a quiescent state. We have performed a detailed time series analysis of the resulting 
light curves. Unusually, we do not detect any orbital modulation in the UV, 
with only marginal evidence for X-ray modulation on this period. Although 
there are peaks in the X-ray periodograms at periods less than 5000 s, 
we attribute them to red noise effects and assign significance to them using 
a novel approach. The variability in the UV and optical bands can also be 
modelled as a result of aperiodic variability  (red noise) in the system. 
There is evidence that  the UV and X-ray fluxes are anti-correlated with a 
time delay of about 100 s, with the UV lagging behind the X-ray emission. 
This anti-correlation is intriguing, but is only present on two occasions lasting 
several 1000 s each. The X-ray spectrum shows  similar emission features to 
other dwarf novae and is well fitted using a multi-temperature emission model. 
We measure a relatively high X-ray luminosity of $\sim1.4\times 10^{32}$ ergs/s, 
although this is consistent with a low binary inclination. Finally, we find evidence 
for a possible -1200km/s blue shift  in the fitted Fe K line energies, possibly indicating
 the presence of an outflow in this low inclination system.

\keywords{accretion, accretion discs -- stars:
individual: YZ Cnc -- novae, cataclysmic variables -- X-rays: stars}}

\maketitle
\section{Introduction}

Dwarf novae are interacting stellar binary systems in which material
gets transfered from a red dwarf secondary star onto a white dwarf via
Roche lobe overflow. In the absence of a strong magnetic field this
material forms an accretion disk around the white dwarf. They show
outbursts which occur on week to month timescales. They have been well
studied in many energy bands, including the optical and X-ray
bands. In X-rays, the emission is suppressed during an outburst while
it only recovers at the end of the optical outburst (eg Wheatley et al
1996). This is due to the gas being optically thick during outburst
but optically thin in quiescence (eg Popham \& Narayan 1995).

YZ Cnc is a dwarf nova with an orbital period of 125.17 min (van
Paradijs et al 1994). It is also a member of the SU UMa class of dwarf
novae - systems which show normal dwarf novae outbursts and also
longer and brighter outbursts. UV observations show strong lines with
P Cygni profiles which vary in shape over the orbital period - this
was interpreted as an asymmetry in the wind flow pattern (Drew \&
Verbunt 1988, Woods et al 1992). White light observations by Pezzuto,
Bernacca \& Stagni (1992) show some evidence for dwarf nova
oscillations on a period of 26 s, although this was only seen on one
night.

\begin{figure*}
\begin{center}
\setlength{\unitlength}{1cm}
\begin{picture}(16,10)
\put(2,9){\includegraphics{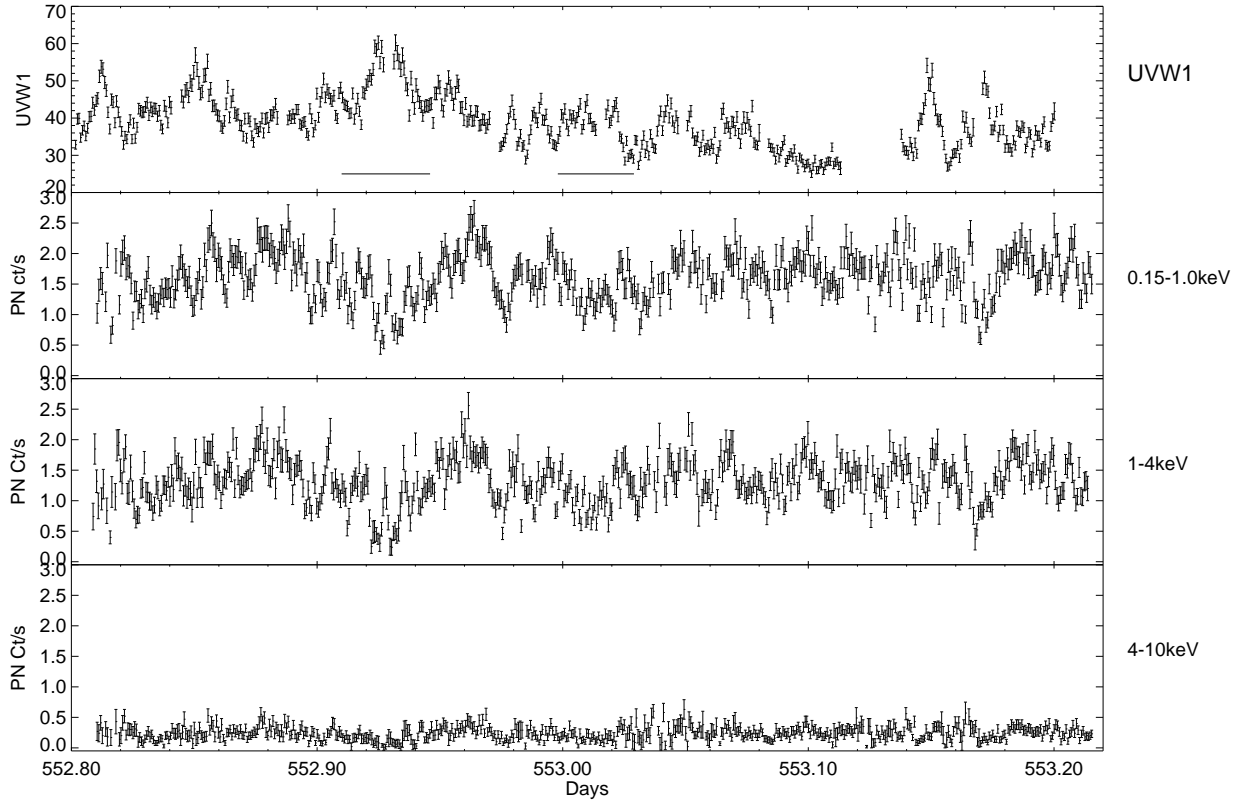}}
\end{picture}
\end{center}
\caption{The light curves of YZ Cnc, from the top; the UV data;
0.15-1.0keV, 1--4keV and 4--10keV. The date is HJD-2452000.0. The bin
size is 60 s. The times of anti-correlation between the X-ray and UV bands
are shown with horizontal lines in the top panel.}
\label{nofold}
\end{figure*}

The most prominent period seen in dwarf novae in the optical band is
the binary orbital period. Other observed periods include the
superhump period (which is due to disc precession), dwarf novae
oscillations (on timescales of several 10's of s) and quasi-periodic
oscillations which have also been seen in X-rays and have modulation
periods of $\sim$100-1000 s (eg Patterson, Robinson \& Nather 1977,
Woudt \& Warner 2002 and references therein).

Unlike the magnetic systems, direct evidence of the spin period of the
white dwarf has been hard to come by.  Recently, observations using
{\xmm} of the dwarf nova OY Car showed the X-rays varied on a
timescale of 2240 s (Ramsay et al 2001a): the amplitude of this
variation was stronger at soft X-rays. A further investigation also
showed a period near 3500 s (Hakala \& Ramsay 2003). They
interpreted the 2240 s period as the spin period of the white dwarf
and the 3500 s period as a beat between the spin and the binary
orbital period.

This led us to investigate whether other dwarf novae, if they are
observed for a sufficiently long time, and with X-ray detectors of
sufficiently large area, show similar evidence for variations which
could be attributed to the spin period of the white dwarf. We have
obtained observations of YZ Cnc using {\xmm} which cover $\sim$4.5
orbital cycles. Our prime goal was to search for modulations in the
X-ray and UV light curves. Here we present the results of our time
series analysis. We also model the X-ray spectra and compare
it with previous X-ray observations of dwarf novae.

\section{Observations}

\subsection{{\xmm} observations}

{\sl XMM-Newton} was launched in Dec 1999 and has the largest
effective area of any imaging X-ray telescope. It has 3 medium
spectral resolution CCD type cameras on-board: two EPIC MOS detectors
(Turner et al 2001) and one EPIC pn detector (St\"uder et al 2001). In
addition it has two high resolution grating spectrographs (RGS), den
Herder et al (2001). Further, it has an 0.3m optical/UV imaging
telescope (the OM) which enables simultaneous X-ray optical/UV
observations (Mason et al 2001). 

YZ Cnc was observed using {\xmm} during 5 Oct 2002: the observation
log is shown in Table 1. The OM was configured in fast mode
and used the UVW1 filter which has an effective wavelength of 2910\AA
(range 2400--3200\AA). The data were processed using the {\xmm}
Science Analysis System v5.4. Data were extracted from an aperture
centered on YZ Cnc and also from source free areas of the detector and
suitably scaled to produce background subtracted light curves and
spectra. Events times were corrected to the arrival at the solar
system barycenter. For lightcurves, events with PATTERN=0-12 were
used, while for spectra we used PATTERN=0-4. We only used events with
FLAG=0  (See for instance XMM-User Guide or XMM- ABC guide 
(heasarc.gsfc.nasa.gov/docs/xmm/abc/) for discussion on different flags
on data). Various (short) time intervals were removed during the
standard processing because of high particle background events. 
We have analysed the observations made using both the EPIC pn and
EPIC MOS detectors. For brevity, in the timing study we report only
the EPIC pn data (the EPIC MOS data is consistent with the results
found from the EPIC pn data), while in the spectral analysis we
utilise all the X-ray data.
The EPIC pn and OM light curves covering the whole observation
are shown in Figure 1.

\begin{table}
\begin{center}
\begin{tabular}{llrr}
\hline
EPIC pn & full-frame & thin & 35506 \\
EPIC MOS & full-frame & thin & 36837 \\
EPIC RGS & full-frame & thin & 36977 \\
OM  &  fast & UVW1 & 29700 \\
\hline
Optical & $V$ & 2002-10-09 & 1hr 50min \\
\hline
\end{tabular}
\end{center}
\label{log}
\caption{The observation log for XMM-Newton observations of YZ Cnc}
\end{table}

We can compare our observed count rate (3.4 ct/s in the 0.15-10keV
EPIC pn energy band) with that seen using {\sl Rosat}, when it was
observed in outburst and quiesence (Verbunt, Wheatley \& Mattei 1999).
We assumed a thermal plasma model and relatively low absorption
($N_{H}=1\times10^{20}$ \pcmsq). Based on the counts observed using
{\sl ROSAT}, and using {\tt PIMMS}, we expect 0.8 ct/s and 3.6 ct/s in
the EPIC pn for YZ Cnc in outburst and quiescence respectively. Our
{\xmm} X-ray observations indicate that YZ Cnc was in a quiesence
state at the time of observation. This is also confirmed from the
optical AAVSO data.

\subsection{Ground based optical observations}

Observations obtained from the AAVSO show YZ Cnc was in outburst
$\sim$2 weeks before the {\xmm} observations. A CCD $V$ band AAVSO
observation taken on the day of the {\xmm} observations show YZ Cnc
was at 15.2 mag - consistent with YZ Cnc being in quiescence.

\begin{figure}
\begin{center}
\setlength{\unitlength}{1cm}
\begin{picture}(6,5)
\put(-0.5,1.5){\includegraphics{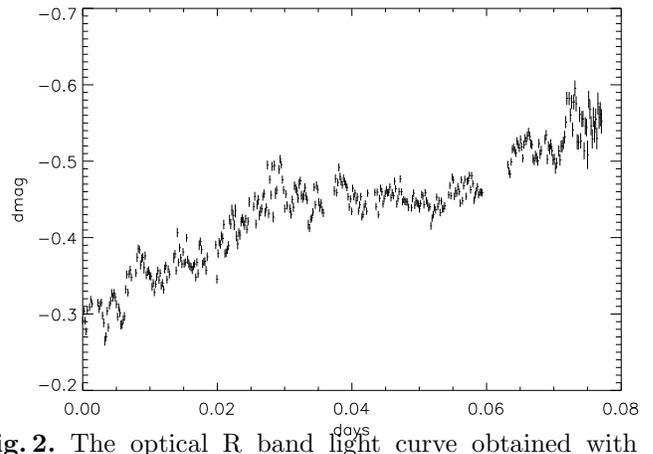}}
\end{picture}
\end{center}
\caption{The optical R band light curve obtained with a 1.2m telescope of
Kryoneri station of the National Observatory of
Athens covering 89\% of the orbital
period.}
\label{optical}
\end{figure}

R-band (Bessell filter) data with a time resolution of 12 s were
obtained on 9 Oct 2002 (4 days after the {\xmm} observations) with the
1.2m telescope at the Kryoneri station of the National Observatory of
Athens. In total, 390 CCD images - covering 0.9 of the binary orbit
were obtained. A comparison star was included in the field of view so
that differential photometry could be obtained. The images were
reduced in the standard manner. The resulting light curve is shown in
Figure \ref{optical}.

Further AAVSO observations show that YZ Cnc started to brighten 1--2
days after the {\xmm} observations and reached V=12.2 on the day of
our R-band observations. The R-band observations were therefore taken
when the system was brightening to outburst. The rise is probably also evident 
in our optical data (Figure 2) , even if it is not possible to properly separate  
the rising trend from the orbital modulation due to the limited time baseline.

\section{Light Curve analysis}

We show the unfolded light curves in Figure \ref{nofold}. The UV data
shows prominent flaring activity, with the `flares' lasting several
100 s. There appears to be modulation on shorter periods as
well. Variability is evident in all X-ray energy bands, and it is more pronounced
in the first half of the observation.

\subsection{Power Spectra}

\begin{figure}
\begin{center}
\setlength{\unitlength}{1cm}
\begin{picture}(6,8)
\put(-0.5,1.8){\includegraphics{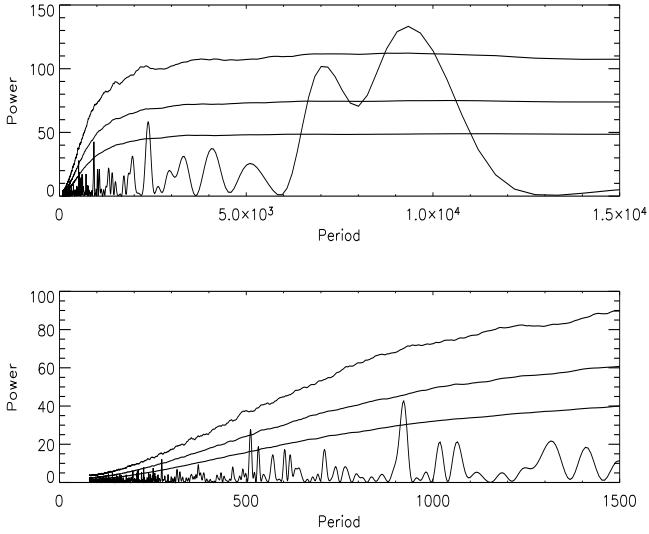}}
\end{picture}
\end{center}
\caption{The power spectrum of the 0.2--4keV EPIC pn light curve.
90\%, 99\% and 99.9\% confidence limits from the Monte Carlo simulations
(including white and fitted red noise) are shown with horizontal lines.
The lower panel is a blowup of the upper panel at shorter periods.}
\label{psd}
\end{figure}

To make a more detailed investigation of the time variability we
obtained a light curve in the 0.2-4keV energy band.  We used the
standard Lomb-Scargle power spectrum analysis. We show the resulting
power spectrum in Figure \ref{psd} (together with significance limits). 
The most prominent peak is at 0.11
days (9300 s). A second broad feature is seen at 7000 s. These
peaks are rather wide due to the 35 ksec length of the
observation. Thus they are consistent with them being due to the
orbital modulation and a beat period between the orbital modulation
and the observation window length. However, we also analysed the light
curves in two equal parts and this analysis showed that these broad
spikes are only present in the first half of the data: this is clear
from Figure 1, where most of the longer term X-ray variability seems
to disappear towards the end of the run. There are also several spikes
at shorter periods, which based on this initial analysis appear
significant, but the confidence limits derived in Sect 3.2  
and plotted in Figure 3 show that these peaks are not significant.

We also performed a similar analysis on the UV and optical data using
the Lomb-Scargle method. The UV data did not show any evidence for the
orbital period so only an analysis of shorter periods is presented here. The optical
data only covers 0.9 of the orbital period, so again only short period
analysis is presented.

We show in Figure \ref{uv-opt} the power spectra in  the optical and UV 
covering the period range up to 3000 s. The strongest spike in the UV periodogram
(top panel) is the spike near 1530 s. From our red noise analysis,
we can however conclude that this spike is less than 99\% significant.
A spike near at a similar period  is seen in the R band (lower panel). 
The fact that the R band spike is much wider is due to the short length of 
the optical observations. The R band power spectrum also shows another, 
stronger spike at about 800 s. This spike is, within the error limits, the 
second harmonic term of the 1530 s period and could possibly appear
in the power spectrum due to a double peaked nature of the optical modulation.
However, also none of the R band spikes exceed the 99.9\% threshold obtained
from the red noise analysis.

\subsection{Red Noise effects}

It is well known that a presence of red noise (noise dominated by low
frequencies) in the data can lead to appearance of spurious spikes in
the power spectra. This 'noise'  is produced by low frequency
aperiodic (random walk type) intrinsic variability in the source (i.e. flickering). 
In order to obtain confidence limits which take into account these
effects in the power spectrum  analysis,  we have proceeded to model the 
X-ray and UV data with time series analysis tools commonly used in analyzing 
evenly sampled time series data in other fields (see for instance Chatfield, 1989) . 
The autoregressive processes used here have also previously been used 
in context of astronomical data analysis (see for instance Roques, Schwarzenberg-Czerny,
Dolez, 2000 and Chen et al. 2000), but not, in our knowledge, for estimating 
the significance of peaks in power spectrum in the presence of red noise.

In order to apply these techniques the data must be evenly spaced in time.
As both X-ray and UV data contain short gaps, we need to interpolate 
flux values for these times (cf. \S2). However, as these gaps are very short 
in time and only account for a very small fraction of the total light curve,
we believe our approach is justified.  

Our approach begins with modelling of the time series as an
Autoregressive (AR) process, where the flux at time $t, F_{t}$ depends on
the fluxes at times $t-1,t-2,...,t-n$ in the following manner:

\begin{eqnarray}
F_{t}  =\sum_{i=1}^{n}{
\alpha_{i}(F_{t-i}-\overline{F})}+Gauss(\overline{F},F_\sigma)
\end{eqnarray}

where $\overline{F}$ is the mean flux, $F_\sigma$ is the standard
deviation of the white noise component and
$Gauss(\overline{F},F_\sigma)$ is a Gaussian random variable with mean
$\overline{F}$ and standard deviation $F_\sigma)$. Depending on the
choice of $n$ we have AR processes of different order (AR(1), AR(2),
..., AR(n)).

Fitting an AR process to the time series is a straight forward linear
least squares problem.  Let {\bf R} be the autocorrelation matrix of a
given time series (up to the lag n-1), and {\bf r$^{T}$} be a vector
of autocorrelations: ${\bf r}^T = (r_1,r_2,...,r_n)$. Then the
$\alpha_i$ coefficients of the AR(n) process can be solved from the
matrix equation:

\begin{eqnarray}
{\bf \alpha = R^{-1}r}
\end{eqnarray}

First tests, using X-ray data, with AR(1) and AR(2) model
fits could not reproduce the original power law slope seen in the X-ray
power spectrum. We then increased the AR process order to 7, which
enabled us to reproduce the power spectrum slope correctly, thus
indicating that the model then contained the right amount of red noise
and other long period aperiodic variability. Actually the last (7th)
coefficient $\alpha_7$ was already an order of magnitude smaller than
$\alpha_6$ indicating that our AR(7) model contained enough orders.
Figure \ref{logpower}. shows the X-ray power spectrum in log(f)-log(Power) space.
Overplotted is our red noise fit using the AR(7) model, that clearly 
models the red noise in the data adequately.

Having reproduced the noise power spectrum satisfactorily, we then
proceeded to create simulated data based on this model. This was done
in the following manner: For every simulated data set we took the
original time points and the AR(7) fit as the basis, but reshuffled
the original data points and added them in random order to the values
generated by the AR(7) process . The reason for using this bootstrap-like 
technique instead of pure gaussian random numbers is that at the observed 
count rate level, the flux distribution is Poissonian, not yet Gaussian. This
process was repeated in order to produce 10000 simulated datasets, all
of which contained all the long term aperiodic variability and red
noise components, as well as the window and sampling effects present
in the data.  These "datasets" could then be used as a basis for
reliable significance level estimates for the Lomb-Scargle power
spectra. We have overplotted the resulting significance limits for X-ray 
data in Figure \ref{psd}.

\begin{figure}
\begin{center}
\setlength{\unitlength}{1cm}
\begin{picture}(6,6)
\put(-1.5,1.0){\includegraphics{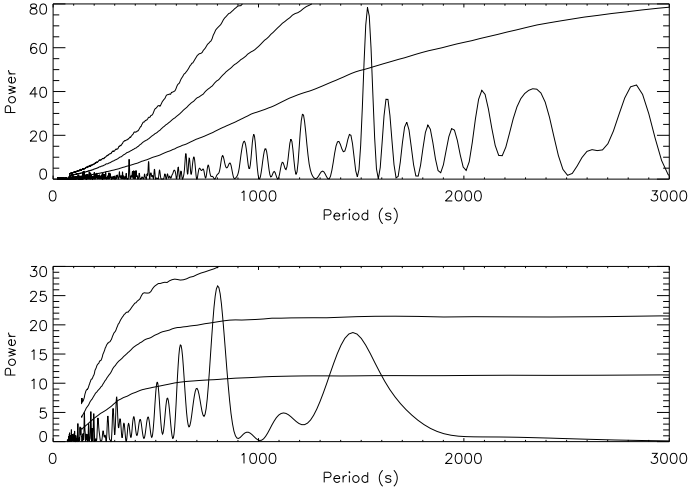}}
\end{picture}
\end{center}
\caption{The UV (top) and R band (bottom) power spectra with 90, 99 and 99.9\% significance levels based on red noise fits. }
\label{uv-opt}
\end{figure}

\begin{figure}
\begin{center}
\setlength{\unitlength}{1cm}
\begin{picture}(6,6)
\put(-1.5,3.5){\includegraphics{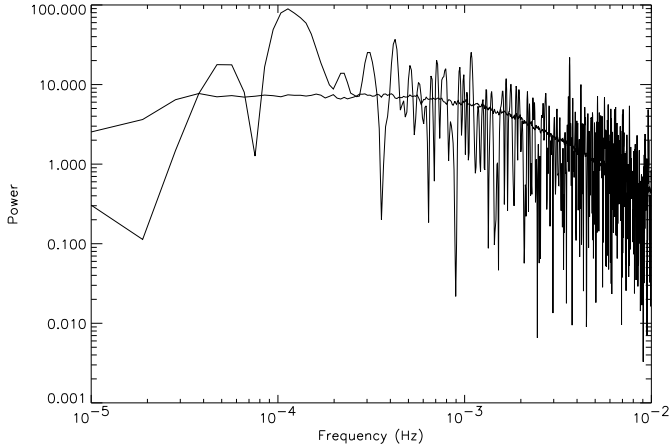}}
\end{picture}
\end{center}
\caption{The observed X-ray power spectrum continuum with an AR(7) red
noise fit .}
\label{logpower}
\end{figure}

Similar analysis was performed on UV and optical data. For the UV 
observations, the data gaps were interpolated, but as the optical data was 
obtained during a continuous, evenly sampled observing run, no interpolation was required. Again, we generated
10000 fake light curves with similar (fitted) red noise properties as in 
the real data. Then the same Lomb-Scargle analysis was repeated for
these fake datasets in order to produce the significance curves overplotted
in Figure \ref{uv-opt}.


\subsection{Significance of  power spectra peaks}

Our autoregression analysis is capable of explaining almost all spikes
in the X-ray power spectrum as a result of red noise. The only
exception is the 9300s spike, which however, as noted earlier, only
appears in the first half of the data.

Similarly, even if clear spikes are seen in the optical and UV light curves,
we can explain them as a result of intrinsic, aperiodic red noise type
variability in the source. No spikes in UV or optical exceed the 99.9\%
(3.3$\sigma$) significance level. Furthemore the 1530s spike in the UV
band is almost entirely dominated by the presence of two strong, flare like, 
events near the end of observation (at around 553.15-553.18, Figure 1.).
If we only analyse the UV light curve before the large gap in data at around
553.12, the spike at 1530 s almost totally disappears from the power spectrum. 
This further strengthens the case for red noise interpretation.

\section{The X-ray - UV cross correlation analysis}

\begin{figure}
\begin{center}
\setlength{\unitlength}{1cm}
\begin{picture}(6,7)
\put(-1.5,3.5){\includegraphics{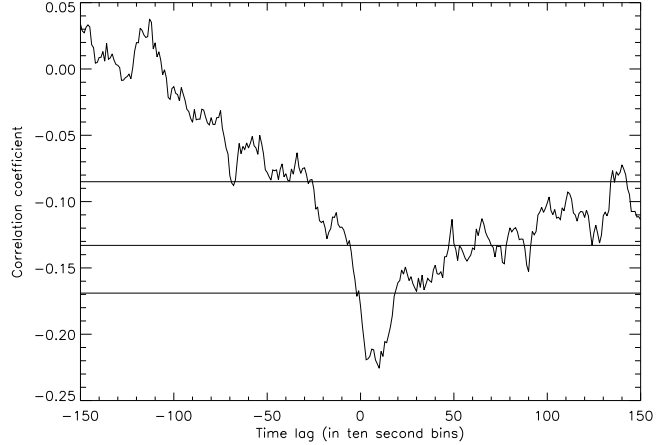}}
\end{picture}
\end{center}
\caption{The average cross correlation between the X-ray and UV bands. We have
overplotted the 90\%(top),99\% and 99.9\% significance limits from our Monte Carlo
simulations.}
\label{ccors}
\end{figure}

There is some suggestion from Figure 1 that there may be an
anti-correlation between the UV and X-ray fluxes (for instance around
552.93 days). We have investigated
this possibility using a cross-correlation analysis. Pandel, Cordova \&
Howell (2003) found that the X-ray and UV fluxes of VW Hyi are
correlated with the X-ray flux following the UV with a 100 s time
lag.

\begin{figure*}
\begin{center}
\setlength{\unitlength}{1cm}
\begin{picture}(18,11)
\put(-1.5,0.5){\includegraphics{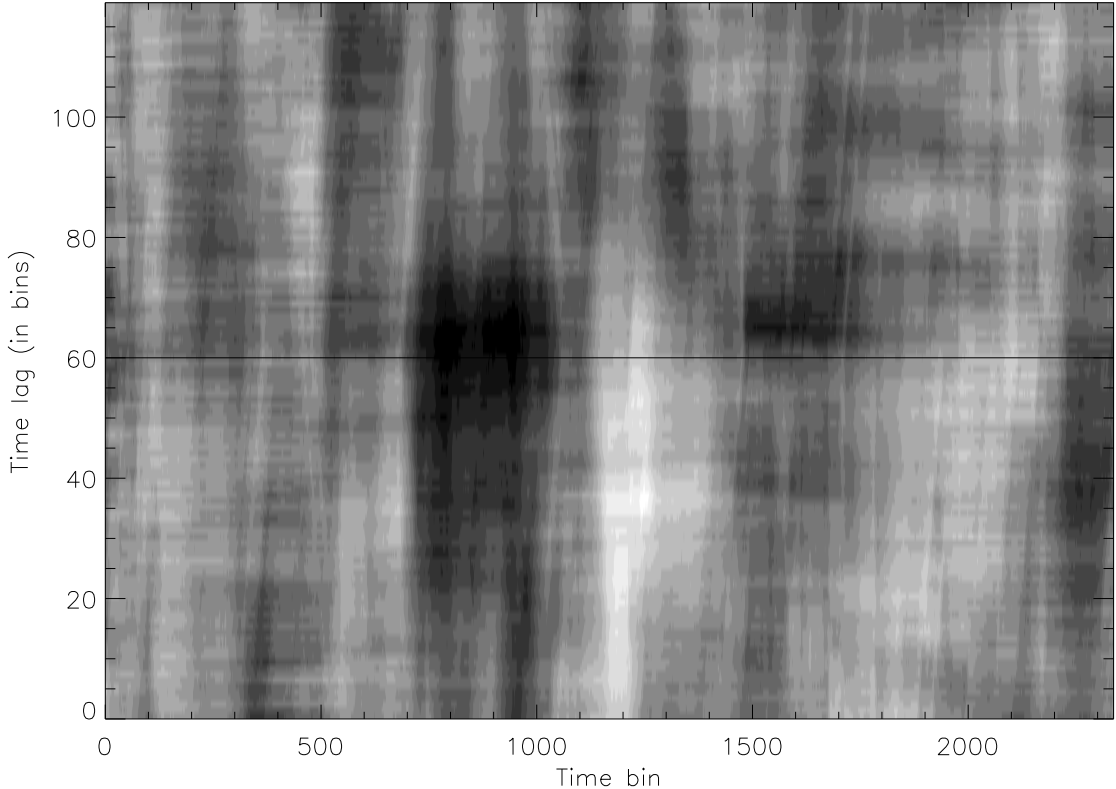}}
\end{picture}
\end{center}
\caption{The time dependent cross-correlation function between the X-ray and UV bands.
The times of strong anticorrelation (darkest parts) are marked by horizontal lines in Figure 1.}
\label{sl-ccors}
\end{figure*}

As the UV and X-ray time points are not exactly same, in order to
compute the cross-correlation function the UV flux was interpolated to
the X-ray time points. The time resolution used for the correlation
analysis was 10 s. In addition to 'normal' cross-correlation analysis
(Figure \ref{ccors}), we computed the cross-correlation function as a function of time
using a sliding box that covered 10\% of the observation length at 
the time. The resulting time dependent cross-correlation function is
plotted in Figure \ref{sl-ccors}. We find that, on average, timelags 
between --1500 to +1500 s show the UV and X-ray flux are more likely to be
anti-correlated than correlated (Figure \ref{ccors}) . However, Figure \ref{sl-ccors}
indicates that the average anti-correlation is dominated by two separate 'events' in the 
time series. During these events the cross-correlation function shows
a clear minimum roughly at +100 s, which corresponds to the
anti-correlated UV emission lagging 100 s behind the X-ray
emission. The actual values for the correlation coefficients are
however rather small. We discuss the physical implications of
this in the Discussion section. 

Attempting to estimate the significance of the possible
anti-correlation with 100 s time lag, we have performed some Monte
Carlo simulations.  Using the fitted red noise models to the both
X-ray and UV light curves we have then generated 30000 pairs of UV and
X-ray light curves, that carry the same amount of red noise as the
observed data. We have then correlated them in order to estimate the
distribution of correlation coefficients. From this exercise we find
that only 3 out of 30000 correlations produce a correlation
coefficient larger (in absolute value) than the value corresponding to
the 100s delay (-0.225). The result was the same both for standard
and rank correlation analysis. The resulting significance levels are
overplotted in Figure \ref{ccors}.
 
However, we have to be  cautious interpreting this result:
Our simulations also show that the distribution of the correlation
coefficients is sensitive to the amount of red noise in the
data. Any underestimation of the amount of red noise in the data would
lead to overestimation of the significance of the correlation.

\begin{figure*}
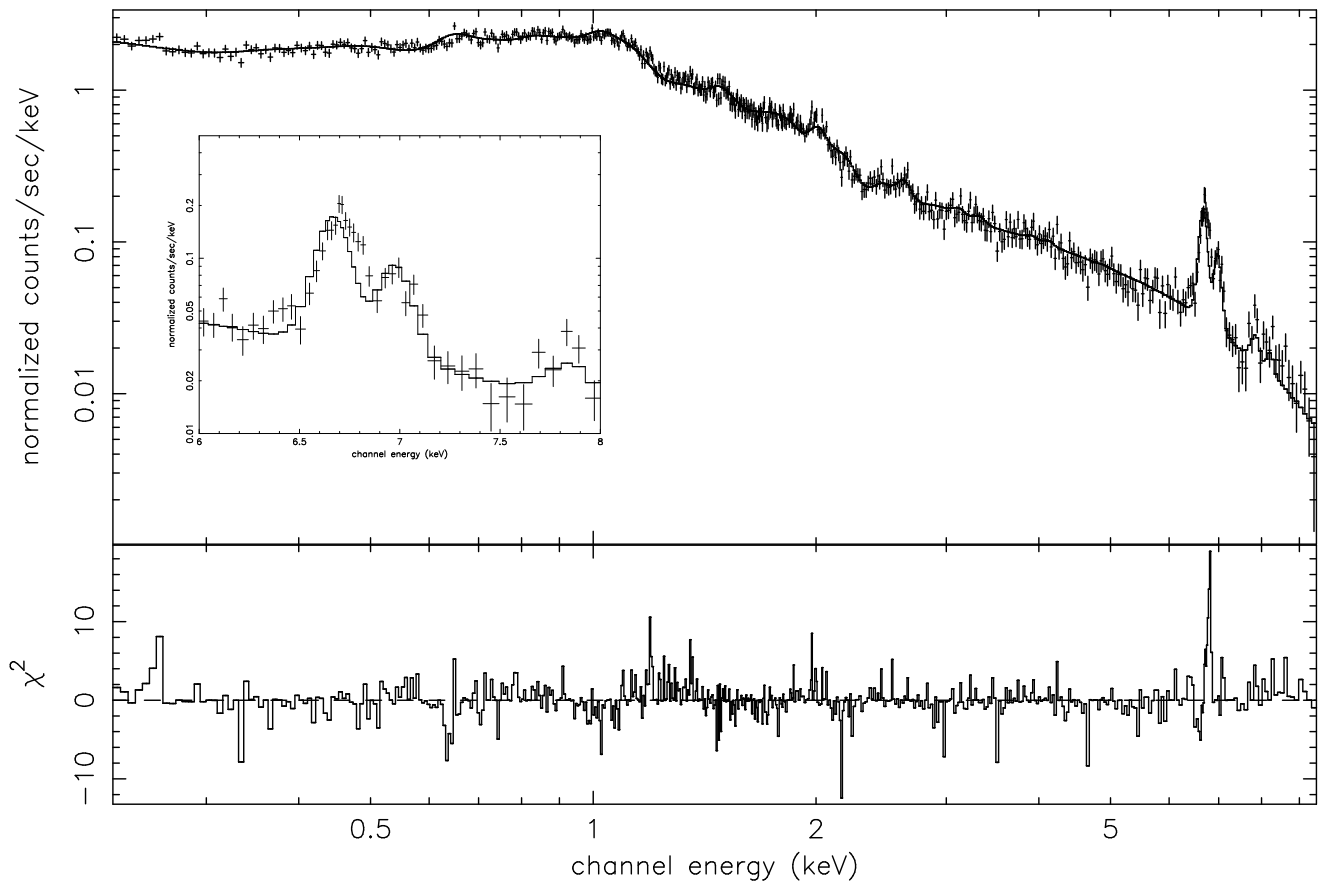

\begin{center}
\setlength{\unitlength}{1cm}
\begin{picture}(18,12)
\put(-1.0,14.5){\includegraphics{spec_paper.ps}}
\put(1.8,10.5){\includegraphics{spec_fe.ps}}
\end{picture}
\end{center}
\caption{The integrated EPIC pn spectrum of YZ Cnc together with a 3-temperature
model fit.}
\label{pn_spec}
\end{figure*}

\def\rchi{{${\chi}_{\nu}^{2}$}}

\section{X-ray spectra}

\subsection{EPIC spectra}

We show the integrated EPIC pn spectrum in Figure
\ref{pn_spec}: its shape is typical of a dwarf nova (eg Baskill,
Wheatley \& Osborne 2004). We fitted this spectrum with an absorbed
(simple neutral absorber) single temperature thermal plasma model:
this gave a poor fit (\rchi=4.95 578dof). We then added a second and
third plasma model, giving improved fits with
\rchi=1.87 (576dof) and \rchi=1.47 (575dof) respectively. We then
allowed the metal abundance to vary from solar, and obtained a fit of
\rchi=1.37 (573dof) with 1.5 solar abundance. We show the spectral 
parameters using this model in Table
\ref{pn_fit} and show the fit in Figure \ref{pn_spec}. There was a small
improvement to the fit (better at the 99.4 per cent level, using an f-test,
\rchi=1.35, 571dof) when we added a neutral absorption model with
partial covering component. We find a very similar fit when we fit the
spectrum with a multi-temperature model with a power law distribution
of emission measures ({\tt cemekl} in {\tt XSPEC}, \rchi=1.37 574dof).
This is similar to the results of {\sl XMM-Newton} observations of the
dwarf novae OY Car Ramsay et al (2001b) and VW Hyi (Pandel, Cordova \&
Howell 2003).

\begin{figure}
\begin{center}
\setlength{\unitlength}{1cm}
\begin{picture}(8,6)
\put(-0.8,7.0){\includegraphics{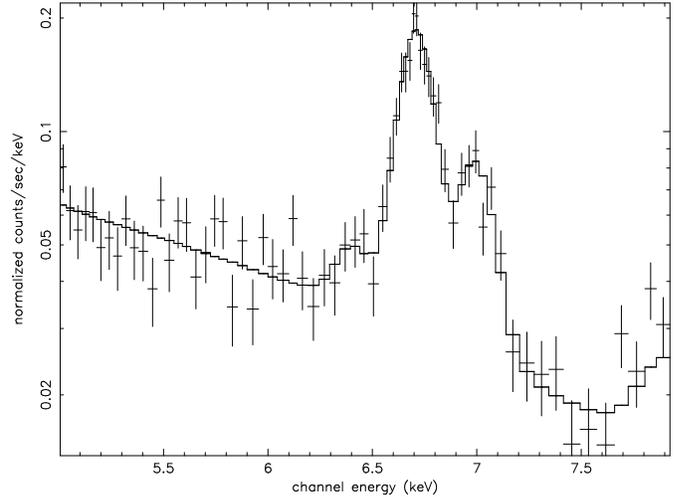}}
\end{picture}
\end{center}
\caption{The fit to the iron line complex with free radial velocity and a 
Gaussian profile at 6.4 keV added. }
\label{spec_blue}
\end{figure}

We also show the residuals to the fit in Figure \ref{pn_spec}: the
residuals are prominent around the Fe K$\alpha$ line. In particular
the 6.7keV line appears at higher energies with respect to the
model. We investigated this further by allowing the redshift to vary
in the fit: we restricted the energy range to 5--8keV. With zero
redshift we obtain a poor fit (\rchi=2.31 63dof), while with a blue
shift of 1200 km/s we get a significantly better fit (\rchi=1.18
62dof). (The confidence interval at the 90 percent level is --1100 to
--1500 km/s). Adding a Gaussian line at 6.4keV to model any fluoresent
line (giving an equivalent width of 35eV, 16-53eV, 90\% confidence limits) 
improves the fit even further (\rchi=1.04 61dof). We show the spectrum together with this
fit in Figure \ref{spec_blue}. We phased events on the orbital period
and extracted four phase-resolved spectra: they all showed blue-shifts
but, within their errors, showed no significant difference in their
velocity shift.

We have considered the possibility that this blue-shift is not
physical and due to instrument calibration effects. We have therefore
examined the combined EPIC MOS data in the same manner as above. The
best-fit blueshift is close to zero, with a 90 percent confidence
interval of --630 to +340 km/s. The difference in the redshift for the
6.7keV line between the EPIC pn and MOS is therefore less than
2$\sigma$. Such 'instrumental blueshift effect' in EPIC pn data has not
been previously reported (Frank Haberl, priv. comm.).

If we {\it do} attribute the blue-shift to a physical process, then
the most likely cause is that we are viewing some kind of outflow,
either wind from the disc or a jet. YZ Cnc has a relatively low inclination
($i\sim30-35^{\circ}$, Shafter \& Hessman 1988): therefore higher
inclination systems might be expected to show lower blue-shifts. We
have re-examined the Fe K$\alpha$ emission line complex of the
eclipsing dwarf nova OY Car ($i=83^{\circ}$). We find that in the
integrated EPIC pn spectrum the best fit red-shift is --220 km/s (the
confidence interval at the 90 percent level is --780 to --170
km/s). This blue-shift is less than that determined in the EPIC pn
observations of YZ Cnc. However, high signal-to-noise X-ray spectra of
dwarf novae with a range of inclination angles is required to
determine if this blue-shift can be attributed to a real physical
affect.

\begin{table}
\begin{center}
\begin{tabular}{lr}
\hline
$N_{H}$ & $8^{+2}_{-0.2}\times10^{19}$ cm$^{-2}$\\
kT (keV) & 0.7$\pm0.05$, 2.0$^{+0.1}_{-0.2}$, 8.3$\pm0.5$\\
Metal Abundance (solar) & 1.5$\pm$0.1\\
Observed flux (0.2--10keV)& $9.24\pm0.4\times10^{-12}$ \ergscm\\
Bolometric flux & $1.21\pm0.05\times10^{-11}$ \ergscm\\
\hline
\end{tabular}
\end{center}
\caption{The fits to the integrated EPIC pn spectra using a
3-temperature thermal plasma model with a neutral absorption model.}
\label{pn_fit}
\end{table}

We note that there is evidence in the EPIC pn data for weak emission
near 6.4keV which might be due to fluorescence from the photosphere of
the white dwarf or cold material.  Ramsay et al (2001b) suggested that
its absence in OY Car was due to the high inclination of that system.
In contrast, YZ Cnc is thought to have a relatively low inclination:
this suggests that inclination does not have a strong influence on
whether the 6.4keV line is detected. This is consistent with the
findings of Baskill, Wheatley \& Osborne (2004) who find that only 4
out of 34 dwarf novae showed a significant (equivalent width of up to
200eV) 6.4keV line in a survey using {\sl ASCA}. Again there was no
correlation with those systems which did show this line and their
binary inclination. It is therefore unclear as to why some systems
show this feature.

Thorstensen (2003) determines the distance to YZ Cnc to be
222$^{+50}_{-42}$ pc from ground based parallax measurements, while
Harrison et al (2003) determined a distance of 320$\pm$40 pc using HST
FGS parallax observations. Thus the X-ray luminosity is likely to be
about $1.4\times10^{32}$ \ergss (assuming 300pc distance). At face
value this is a rather high value for a quiescent dwarf nova. However,
YZ Cnc has a relatively low inclination and the high X-ray
luminosity is then consistent with van Teeseling, Beuermann \& Verbunt
(1996) who found that the observed flux is anti-correlated with
inclination and thus in high inclination systems a significant
fraction of the X-rays from the boundary layer are probably absorbed
by the accretion disk. Furthermore, at least two systems (V603 Aql and 
SS Cyg) in Baskill et al. (2003) have luminosities similar or higher
than YZ  Cnc. 

\subsection{RGS spectra}

\begin{figure}
\begin{center}
\setlength{\unitlength}{1cm}
\begin{picture}(6,6.5)
\put(-0.,8){\includegraphics{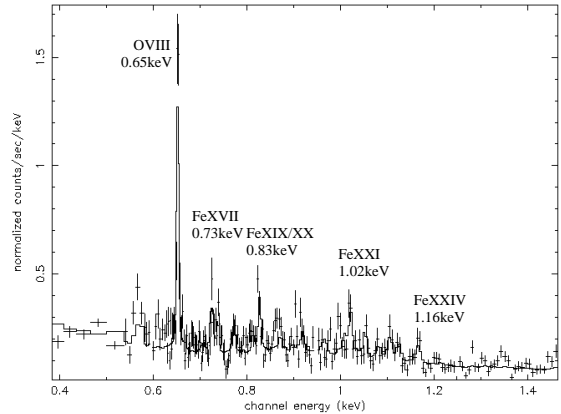}}
\end{picture}
\end{center}
\caption{The integrated RGS spectrum of YZ Cnc with the
multi-temperature best model fit to the EPIC pn data shown as a solid
line.}
\label{rgs_spec}
\end{figure}

We show the integrated RGS(1+2) spectrum in Figure \ref{rgs_spec}. The
most prominent feature is an emission line at 0.65keV which is due to
OVIII Lyman$\alpha$. Similar to VW Hyi (which also shows this
prominent line), it is slightly broadend compared to the instrumental
response (590 km/s, 120-770 km/s, 90\% confidence). We show the spectrum
compared with the best fit model obtained from the fits to EPIC pn spectrum. 
Although the signal to noise is relatively low there is evidence for other emission lines
including Fe XVII (0.73keV), Fe XIX,XX (0.73keV), Fe XXI (1.02keV) and
Fe XXIV (1.16keV). Such a range of charge states indicates a range of
temperature regions, which is also implied by the need for
multi-temperature spectral fits for the EPIC data.

\section{Discussion}

The main aim of this programme was to obtain information about the
temporal behaviour of YZ Cnc. In particular, the $\xmm$
observations of a similar system OY Car (Ramsay et al. 2001a, Hakala
\& Ramsay, 2003) have indicated that the X-ray light curve could
contain a significant white dwarf spin period component.

Our analysis of the EPIC and OM data has revealed intrinsic variability
in the source, all of which can be attributed to the effects of red
noise. Baskill et al. (2003), have analyzed a complete set of ASCA observations
of non-magnetic CV's. They conclude that 7 out of 34 systems they have 
analyzed show evidence for X-ray modulation at the orbital period. Most of these
systems, however, are high inclination systems in quiescent state. 
Thus the fact that we do not detect any persistent orbital modulation in
our X-ray light curve is in agreement with the relatively low inclination of
YZ Cnc.
 
According to our study, all the spikes seen in the X-ray power spectrum
are due to the red noise. There is no sign for the previously reported 26s
period that was earlier seen in the optical white light data during
one single night (Pezzuto, Bernacca \& Stagni, 1992) .  

Curiously, also our OM UV data does not show any periodic modulation at 
all. Based on results of other dwarf novae (for instance OY Car, Ramsay et al., 2001a
and VW Hyi, Pandel et al., 2003) an UV modulation due
to the hot spot on the edge of the disc would be expected. However, recently
Hakala \& Ramsay (2003) showed using $\xmm$ OM data that even in a 
high inclination system like OY Car, the hot spot modulation in quiescence 
is much more prominent in optical band than in UV. This, combined with 
the presumably low inclination, is the likely reason for nondetection. 

Our cross-correlation analysis analysis has revealed that the X-ray and UV fluxes
seem to be occasionally anti-correlated. This anti-correlation seems to be related
to the times of UV flares as indicated in Figures \ref{nofold} and \ref{sl-ccors}. When
the anti-correlation is present, the UV flares seems to lag behind the drop in X-ray
emission by  $\sim$ 100 s. This is very curious: Lags at similar time scales have
 been observed in VW Hyi (Pandel et al. 2003), but in their case the X-rays and UV 
are correlated and the X-rays lag behind the UV. Such correlation arises naturally
if the UV emission originates near the transition region from disc to the boundary layer
and the propagation time of accretion rate fluctuations from the UV emitting region
to the boundary layer (X-ray source) is of the order of 100s. 
Interestingly  Pandel et al. (2003) also note that they see X-rays and UV vary strongly 
at time scales of 1500s, which also seems to be the case for UV emission in YZ Cnc.

Given that the variability time scale is the same, but the cross-correlation is opposite
we have to conclude that the observed difference is most likely due to different viewing
geometry (inclination) of the system. This is also suggested by the lack of orbital modulation
in UV. However, the exact accretion geometry and/or physics that would produce such an effect
as a function of inclination remains unresolved.

\section{Conclusions}

Our $\xmm$ observations of YZ Cnc do not reveal any periodicities
in the system, even if clear short term variability is present in both X-ray
and UV bands. The cross-correlation analysis reveals that during major
UV flares the X-rays and UV are anticorrelated, with UV lagging behind
the X-rays by $\sim$ 100 s.   
 The measured X-ray luminosity, $\sim 1.4\times 10^{32}$
\ergss, is relatively high for a dwarf nova in quiescent state.
A multitemperature plasma model (possibly with a partially covering absorber)
is required to model the X-ray spectrum. Finally we find that the iron line
complex near 6.7 keV shows some evidence for blue shifted emission at
about -1200 km/s. However, we cannot entirely rule out instrumental effects.

\begin{acknowledgements}
This paper is based on observations obtained with XMM-Newton, an ESA
science mission with instruments and contributions directly funded by
ESA Member States and the USA (NASA). PJH is an Academy of Finland
research fellow. We acknowledge with thanks the variable star
observations from the AAVSO International Database contributed by
observers worldwide and used in this research.

\end{acknowledgements}

\end{document}